\documentclass[a4paper]{article}
\usepackage{ctable}
\usepackage{amssymb}
\usepackage{float}
\usepackage{caption}
\usepackage{subcaption}
\usepackage{nicefrac}
\usepackage{amsmath}
\usepackage{graphicx}
\usepackage{multicol,multirow}

\usepackage{tabularx}
\usepackage{lipsum}
\newcommand{\smallsim}{\smallsym{\mathrel}{\sim}}

\usepackage{cite,url}
\usepackage{hyperref}
\hypersetup{
    colorlinks=true,
    linkcolor=purple,
    filecolor=magenta,      
    urlcolor=purple,
}

\makeatletter
\newcommand{\smallsym}[2]{#1{\mathpalette\make@small@sym{#2}}}
\newcommand{\make@small@sym}[2]{%
  \vcenter{\hbox{$\m@th\downgrade@style#1#2$}}%
}
\newcommand{\downgrade@style}[1]{%
  \ifx#1\displaystyle\scriptstyle\else
    \ifx#1\textstyle\scriptstyle\else
      \scriptscriptstyle
  \fi\fi
}
\makeatother

\newcolumntype{Y}{>{\centering\arraybackslash}X}

\usepackage{INTERSPEECH2022}

\title{On The Potential of Jointly-Optimised Solutions to \\ Spoofing Attack Detection and Automatic Speaker Verification}

\name{Wanying Ge$^{*}$, Hemlata Tak$^{*}$\thanks{$^{*}$ These authors contributed equally to this work.}, Massimiliano Todisco and Nicholas Evans}

\address{EURECOM, Sophia Antipolis, France}
\email{lastname@eurecom.fr}

\begin{document}
\maketitle
\begin{abstract}
    The spoofing-aware speaker verification (SASV) challenge was designed to promote the study of jointly-optimised solutions to accomplish the traditionally separately-optimised tasks of spoofing detection and speaker verification. Jointly-optimised systems have the potential to operate in synergy as a better performing solution to the single task of reliable speaker verification. However, none of the 23 submissions to SASV 2022 are jointly optimised. We have hence sought to determine why separately-optimised sub-systems perform best or why joint optimisation was not successful. Experiments reported in this paper show that joint optimisation is successful in improving robustness to spoofing but that it degrades speaker verification performance. The findings suggest that spoofing detection and speaker verification sub-systems should be optimised jointly in a manner which reflects the differences in how information provided by each sub-system is complementary to that provided by the other. Progress will also likely depend upon the collection of data from a larger number of speakers.
\end{abstract}

\section{Introduction}
\label{sec:intro}

    Solutions to reliable biometric voice recognition often comprise a pair of independent sub-systems, namely separately-optimised automatic speaker verification (ASV) and spoofing countermeasure (CM) classifiers. The role of the ASV sub-system is to verify whether or not the voice in test and enrolment utterances corresponds to the same speaker. That of the CM is to differentiate between bona fide and manipulated or artificially fabricated utterances and hence to protect the ASV from being spoofed by an adversary.

    ASV and CM systems are generally learned using different databases. State-of-the-art ASV systems~\cite{desplanques2020ecapa,jung20rawnet2,chung2020in,bai2021sr_overview} are typically trained to capture speaker-discriminative characteristics from massive datasets containing a huge number of utterances captured from a large number of speakers~\cite{mclaren16sitw,voxceleb1,voxceleb2, greenberg2020nist_ser, sadjadi2021nist}. ASV databases contain bona fide speech data. Those used for the development of CMs necessarily contain utterances of both bona fide and spoofed speech. To help marginalise speaker-related influences, they too are collected from a large number of speakers, though usually far fewer than ASV. To promote the learning of generalised countermeasures which perform reliably in the face of previously unseen attacks, CM databases should contain spoofed data generated with a diverse range of spoofing attack algorithms~\cite{todisco2019asvspoof, ASV2021challenge, yi2022add}.  With two separate sub-systems, each designed to solve a different, specific problem, there is then the question of how the two sub-systems function in tandem together to solve the single task of reliable ASV.

    CMs have the potential both to reject spoofed utterances, but also to falsely reject bona fide utterances.  CMs can hence impact upon ASV performance. 
    While not its role, the ASV sub-system also has potential to reject spoofing attacks, e.g.\ 
    the ASV sub-system can decide in favour of a non-target hypothesis in the case that a spoofed test utterance does not reflect well the characteristics of the targeted speaker. The influences of the two sub-systems are hence not independent, implying a potential benefit for their joint optimisation.

    In contrast to the fusion of separately trained ASV and CM systems, joint optimisation has the potential to exploit the synergy between each sub-system so that the strengths of one can compensate for the weaknesses of the other. By way of an example, the optimisation of CMs to reject poorer quality spoofing attacks that will in any case be rejected by the ASV sub-system might be a needless waste of discrimination capacity. 
    Instead, CM optimisation should prioritise the reliable rejection of more potent attacks that would otherwise be successful in spoofing the ASV sub-system~\cite{wang2020asvspoof}.
    
    The Spoofing-Aware Speaker Verification (SASV) challenge~\cite{jung2022sasv,jungsasv2022} was designed, in part, to promote the study of jointly-optimised ASV and CM solutions to protect against text-to-speech (TTS) and voice conversion (VC) spoofing attacks. 
    The work described in this paper aims to shed light on why, despite the apparent merits, joint-optimisation was absent from submissions to the inaugural edition; all of the top-performing systems combine separately optimised sub-systems which are fused at either score~\cite{Alenin2022A,dku2022sasv,Zhang2022Norm,heo2022IRLab} or embedding~\cite{hyu_sasv_2022,zhang2022flyspeech, heo2022IRLab} levels. Perhaps joint-optimisation was unsuccessful, or simply not explored. We aim to show whether the SASV task is best solved using separately optimised CM and ASV sub-systems, or if joint-optimisation merits attention in the future.

\section{Related work}
\label{sec:relatedwork}

    Two different, general approaches can be found among the submissions to the inaugural SASV challenge.
    The first involves the fusion of separately-optimised, i.e.\ pre-trained and fixed CM and ASV sub-systems at decision, score or embedding levels. Decision-level fusion can be realised from the cascaded combination of ASV and CM sub-systems such that decisions are combined in the form of a logical-AND operation~\cite{sahidullah2016integrated}. Such systems accept only utterances that are labelled by the CM as bona fide and by the ASV system as a target trial; all others are rejected. CM and ASV sub-systems can also be combined in parallel using score\cite{todisco2018integrated} or embedding~\cite{khoury2014introducing} level fusion or classification. In contrast to cascaded combination, parallel combination exploits information provided by both sub-systems for every single utterance. Fusion and classification are efficient approaches to combine pre-trained, fixed CM and ASV sub-systems as a solution to the SASV task and typify all of the top-performing submissions to the first SASV challenge~\cite{Alenin2022A,dku2022sasv,Zhang2022Norm,heo2022IRLab, hyu_sasv_2022,zhang2022flyspeech}.
    
    \newpage

    More closely integrated systems have also been explored.  These solutions discriminate between bona fide target trials and non-target or spoofed trials using a single model or classifier. Such SASV solutions involve either the fine-tuning of pre-trained ASV systems~\cite{kang22sasv} to improve robustness to spoofing, or multi-task learning (MTL) techniques~\cite{caruana1997multitask} in which 
    spoofing detection and speaker verification functionalities are learned using composite loss functions~\cite{li2019multi,li2020joint,shim2020integrated, teng2022Magnum, zhang2022hccl, ta2022vtcc}. While they can be more efficient, single-classifier systems tend to be outperformed by solutions that combine separately-optimised CM and ASV sub-systems.
        
\section{Optimisation framework}
\label{sec:proposed}

\begin{figure}[!t]
         \centering
         \includegraphics[trim=6cm 0 3cm 0,clip,scale=0.5]{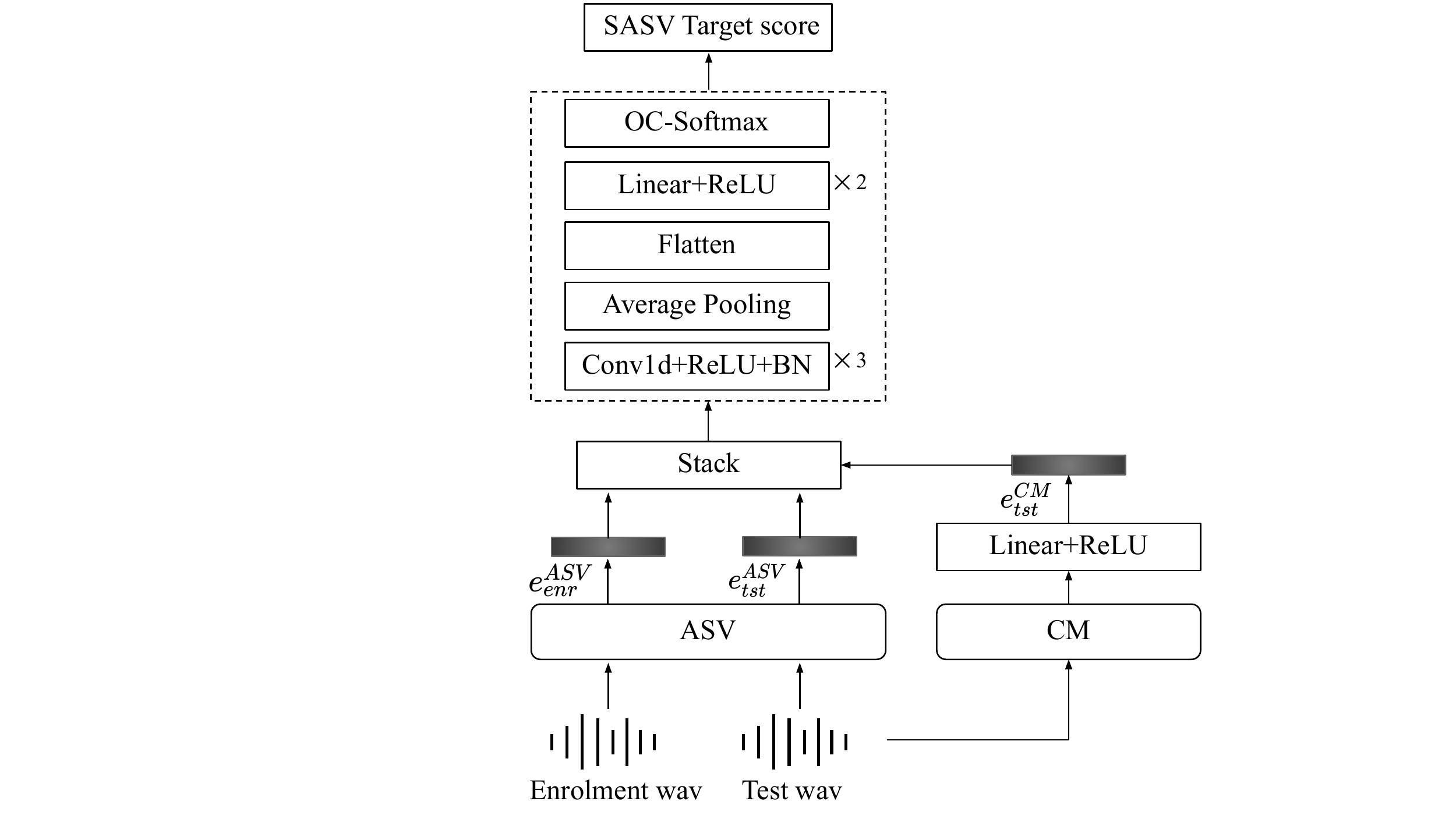}
        \caption{Framework for separate and joint optimisation.}
        \label{fig:sasv}
        \end{figure}

    The meaningful comparison of SASV solutions comprising CM and ASV sub-systems that are optimised either separately or jointly depends upon the use of an architecture in which all other components are common to both approaches.  The architecture designed specifically for this work is illustrated in Fig.~\ref{fig:sasv}. It operates upon stacked embeddings extracted from an ASV sub-system (bottom left) and a CM sub-system (bottom right). These are fed to a back-end classifier (top left), the output of which is an SASV score for the bona fide target class.  

    \subsection{ASV sub-system}
    \label{sec:asv}
    
    The ASV sub-system is the model described in~\cite{heo2020clova}, namely a ResNet34 model with squeeze-and-excitation (SE) blocks~\cite{hu2018squeeze} (ResNetSE34). 
    Compared to all other tested alternatives, including an ECAPA-TDDN system~\cite{ecapatdnn_pretrained}, we found this particular system to give better results when combined with a spoofing detection system.
    A feature extraction layer is first used to decompose input waveforms into spectro-temporal representations. 
    Four convolutional layers with SE blocks are then used to extract compact, deep features. After a flattening operation, attentive statistics pooling (ASP)~\cite{okabe2018asp} is used to capture long-term speaker characteristics and to project the variable length input to a fixed-length embedding vector. Embeddings are extracted in the same way for both the enrolment utterance $e^{ASV}_{enr}$ and the test utterance $e^{ASV}_{tst}$.
    The network is optimised through a combination of softmax and angular prototypical loss~\cite{wang2017angularloss}.

    \subsection{CM sub-system}
    \label{sec:cm}
    The CM sub-system is the state-of-the-art SASV 2022 baseline CM named AASIST~\cite{jung2022aasist}. AASIST uses a RawNet2-based encoder~\cite{jung2020improved,tak2021rawnet2} to extract high-dimensional spectro-temporal features directly from waveform inputs. A spectro-temporal graph attention network (RawGAT-ST)~\cite{tak2021end} with heterogeneous graph attention layers and max graph operations are then used to integrate temporal and spectral representations. CM output scores are generated using a readout operation and a hidden linear output layer comprising two classes, with CM embeddings $e^{CM}_{tst}$ being extracted from the penultimate layer. The AASIST model is trained with a weighted cross-entropy loss function.
    
    \subsection{Back-end classifier}
    A convolution neural network with adaptive average pooling is used as a back-end classifier to learn discriminative information from ASV and CM embeddings~\cite{zhang2022flyspeech}. The lower-dimension CM embedding $e^{CM}_{tst}$ extracted from the test utterance is first transformed using a linear layer to have the same dimension as the pair of ASV embeddings $e^{ASV}_{enr}$ and $e^{ASV}_{tst}$. The extracted embeddings are stacked along a new dimension. Three 1D convolutional layers are used to expand this dimension and to capture the variance between enrolment and test ASV embeddings, in addition to that between ASV and CM embeddings. 1D adaptive average pooling is then used to aggregate information from the deep representation. Using a one class (OC) softmax loss function~\cite{zhang2021oneclass}, we output one final score to reflect support for the positive bona fide, target class, as opposed to the negative spoofed target and non-target bona fide classes. 
    Aggregated information is mapped to the score using two linear layers 
    and one OC softmax layer such that higher values indicate support for the bona fide target class.

\section{Experimental setup}
\label{sec:experiments}

    Described in this section are the database, protocols, evaluation metrics and implementations used in this work.
    
    \subsection{Databases and protocols}
    \label{sec:database}
    The ASV sub-system (ResNetSE34, Sec.~\ref{sec:asv}) is trained using the development partition of the VoxCeleb2 database~\cite{voxceleb2}. 
    Data augmentation is applied using additive noise recordings in the MUSAN corpus~\cite{musan} and simulated room impulse response (RIR) filters~\cite{rir}. The VoxCeleb1~\cite{voxceleb1} test partition is used to select the best ASV model. 
    The CM sub-system (AASIST, Sec.~\ref{sec:cm}) is trained using the training partition of the ASVspoof 2019 logical access (LA) database~\cite{wang2020asvspoof}. The development partition is used to select the best CM model. The SASV system is trained using the same training partition, but with specific SASV rather than CM protocols.

    All experiments reported in this paper were performed using the SASV 2022 protocols\footnote{\url{https://github.com/sasv-challenge/SASVC2022\_Baseline/tree/main/protocols}}~\cite{jung2022sasv,jungsasv2022}. SASV protocols involve three trial classes: 
    (i)~target, bona fide trials uttered by the same speaker as that of the enrolment utterance(s); 
    (ii)~bona fide, non-target trials uttered by a different speaker as that of the enrolment utterance(s); 
    (iii)~spoofed, non-target trials containing speech which is either synthesised or converted in order to resemble the voice of the speaker of the enrolment utterance(s).

    \begin{table}[!t]
  \caption{Trial types used for estimation of each EER. ``+'' indicates the
positive class and ``-'' indicates the negative class.}
  \centering
  \label{tab:eer_types}
  \begin{tabularx}{\linewidth}{lYYY}
    \hline
    & \textbf{Bona fide} & \textbf{Bona fide} & \textbf{Spoofed}\\
    &\textbf{target}&\textbf{non-target}&\textbf{non-target}\\
    \hline
    SV-EER & + & - &  \\
    SPF-EER & + &  & - \\
    SASV-EER & + & - & - \\
    \hline
  \end{tabularx}
  \vspace{-15pt}
\end{table}
    
    \subsection{Metrics}
    \label{sec:metric}
    As depicted in Tab.~\ref{tab:eer_types}, there are three evaluation metrics, all equal error rate (EER) estimates. 
    The speaker verification EER (SV-EER) is estimated using target and bona fide non-target trials and reflects a traditional approach to estimate ASV performance in the absence of spoofing attacks. 
    The spoofing EER (SPF-EER) is estimated using target and spoofed non-target trials and reflects the vulnerability of the ASV system when non-target trials are replaced with spoofing attacks.
    The spoofing-aware speaker verification EER (SASV-EER) is estimated using trials involving all three classes: target, bona fide non-target and spoofed non-target.
    Tab.~\ref{tab:eer_types} shows that all three metrics estimate the performance of binary classifiers in which the positive class consists in bona fide target trials only whereas the negative class is different for each metric.
    The SASV-EER is estimated with the union of the negative class trials used to estimate the SV-EER and SPF-EER. 

    \subsection{Implementation details}
    \label{sec:implement}

    The ASV sub-system\footnote{\url{https://github.com/clovaai/voxceleb_trainer}} is pre-trained using a random selection of 2-second waveforms extracted from the VoxCeleb2 development set. Waveforms are first pre-emphasised before the extraction of 64-dimensional log Mel-filterbank features using a hamming window of 25~ms with a 10~ms frame shift. Discriminative speaker information extracted from the input feature is then compacted into a 512-dimensional embedding. Inputs to the CM sub-system\footnote{\url{https://github.com/clovaai/aasist}} are raw waveforms of $\smallsim$4 seconds duration (64600 samples). The dimension of the extracted CM embedding is set to 160. Further details can be found in~\cite{heo2020clova} and~\cite{jung2022aasist}.

    The SASV system is trained using enrolment and test utterances both of $\smallsim$4 seconds duration (64600 samples). Stacked embeddings of dimension [3,512] are then expanded to 64 channels, then 128, and finally 256 channels using 1D convolutional layers. A 1D adaptive average pooling layer is then applied to aggregate the stacked, deep representation of dimension [256,512] to dimension [256,4]. After flattening to dimension [1024,1], a pair of linear layers are used to reduce the dimension to [512,1] and then [256,1]. An OC-Softmax layer~\cite{zhang2021oneclass} is then applied to map the 256-dimensional representation to a single score using a scale factor of 10.
     Empirically optimised margins are 0.8 for the positive class (target bona fide) and 0.2 for the negative class (spoofed target and non-target bona fide). 
     
    \subsection{SASV training}
    The SASV system is trained in the same manner as the SASV Baseline~2 solution~\cite{shim2022baseline} but we make modest changes to the inference phases. Instead of extracting embeddings from each utterance and then averaging, which is more computationally demanding, a single ASV enrolment embedding is extracted from the concatenation of the set of enrolment utterances for each speaker. This approach accelerates the inferences processes which then involve the extraction of comparatively fewer embeddings. Concatenated enrolment utterances are fixed to a duration of 60~seconds for the development set and to 90 seconds for the evaluation set (only because they tend to be of greater duration).  Impacts upon performance using concatenated enrolment utterances were found to be negligible.

    Both separately-optimised (pre-trained, then fixed) and jointly-optimised systems are trained with the same database.  
    They share the same, common architecture illustrated in Fig.~\ref{fig:sasv} and the same training hyper-parameters.
    Only under joint optimisation are all network parameters updated during training, including those for CM and ASV sub-systems. 
    Otherwise, only those of the back-end classifier are updated, with fixed CM an ASV parameters. 
    In all cases, trainable parameters are updated for 20 epochs with a batch size of 20 and an initial learning rate of 5e-5. 
    The learning rate is decayed by a factor of 0.95 for every 200 mini-batches following an exponential scheduler. 
    The best model is selected as that which gives the lowest SASV-EER for the development set.
     
    All models are trained once with the same random seed on a single NVIDIA GeForce RTX 3090 GPU. Results are reproducible with the same random seed and GPU environment using the implementation available online.\footnote{\url{https://github.com/eurecom-asp/sasv-joint-optimisation}}

\section{Results}
\label{sec:results}

    \begin{table}[!t]
        \caption{Results for pre-trained, jointly-optimised and baseline systems for SASV 2022 development and evaluation partitions.}
        \vspace{2mm}
        \centerline{
        \resizebox{\columnwidth}{!}{%
        \renewcommand{\arraystretch}{1.3}
        \begin{tabular}{lcccccc}
        \hline
        \multirow{2}{*}{\textbf{System}} & \multicolumn{2}{c}{\textbf{SASV-EER}}  & \multicolumn{2}{c}{\textbf{SPF-EER}} & \multicolumn{2}{c}{\textbf{SV-EER}}\\ 
         &
         \multicolumn{1}{c}{\texttt{dev}}  & \multicolumn{1}{c}{\texttt{eval}} &
         \multicolumn{1}{c}{\texttt{dev}}  & \multicolumn{1}{c}{\texttt{eval}} &
         \multicolumn{1}{c}{\texttt{dev}}  & \multicolumn{1}{c}{\texttt{eval}}\\ 
        \hline

        \textbf{Pre-trained, fixed}  &0.81 &1.15 &0.54 &1.12 &1.73 &1.38\\
        \textbf{Joint optimisation} &1.15 &1.53 &0.27 &0.75 &2.15 &2.44\\
        Baseline1-v2~\cite{jungsasv2022} &1.01 &1.71 &0.23 &1.76 &1.99 &1.66\\
        Baseline2~\cite{shim2022baseline} &4.85 &6.37 &0.13 &0.78 &12.87 &11.48\\
        
        \hline
        \end{tabular}}}
        \label{tab:proposed_systems}
        \end{table}  
        
    Results in columns 2 and 3 of Tab.~\ref{tab:proposed_systems} show SASV-EERs for the SASV 2022 development and evaluation partitions. 
    The pre-trained, fixed system gives the lowest SASV-EERs of 0.81\% for the development partition and 1.15\% for the evaluation partition, whereas those for the jointly-optimised system are 1.15\% and 1.53\% respectively. 
    While both systems outperform both baselines, these results rebut the initial hypothesis that joint optimisation stands to better exploit the complementarity between CM and ASV sub-systems; at the SASV-level, better results are achieved using pre-trained, fixed systems.
    
    Corresponding spoofing detection results in terms of SPF-EERs shown in columns 4 and 5 of Tab.~\ref{tab:proposed_systems} paint a more favourable picture. Results for joint optimisation of 0.27\% and 0.75\% for the development and evaluation partitions respectively, are substantially better than those for the pre-trained, fixed system of 0.54\% and 1.12\%. Lower SPF-EERs suggests that the ASV-system is complementary to the CM sub-system in terms of spoofing detection performance. Of course, this finding might just be the result of fusing two different sub-systems. It might also suggest that use of information in \emph{both} test and enrolment utterances contributes to more reliable spoofing detection performance. This in turn suggests a level of speaker-dependence in the cues used to detect spoofing.By acting to marginalise speaker-related artefacts while emphasizing spoofing-related artefacts, access to both the test utterance, which is either bona fide or spoofed, in addition to the enrolment utterance, which is always bona fide, might also account for improvements in performance.
    
    Speaker verification results in terms of the SV-EER are shown in the two right-most columns of Tab.~\ref{tab:proposed_systems}. The results show why, despite improved SPF-EERs, SASV-EERs degrade. SV-EERs under joint optimisation of 2.15\% and 2.44\% are substantially higher than those for the pre-trained, fixed system of 1.73\% and 1.38\%. From one angle, the finding that the CM does not contribute to improved ASV performance is unsurprising. The CM has access only to the test utterance and not the enrolment utterance; no new speaker-related information is provided by combining CM and ASV embeddings. This finding, though, does not explain why performance degrades.

    \section{Discussion and further analysis}
    
    The results show that joint optimisation succeeds in improving robustness to spoofing, but fails to improve overall reliability.
    This may simply be the result of training a more complex system in which all three components (CM, ASV sub-systems and the back-end classifier) are all trained using the same data that is otherwise used to train each sub-system separately.

    Our initial suspicion was that over-fitting is caused by the lack of speakers in the SASV training dataset.
    Among the SASV 2022 submissions, we noticed one~\cite{zhang2022hccl} that uses a pre-trained, fixed ASV sub-system without fine tuning, despite fine tuning being applied to the CM sub-system.
    This is likely because fine tuning of the ASV sub-systems was not beneficial.
    ASV sub-systems are typically pre-trained using data collected from thousands of speakers, whereas the SASV 2022 training data is collected from only 20 speakers.
    The use of such speaker-sparse training data for fine tuning hence stands to reduce generalisation, i.e.\ increase over-fitting.
    
    Indications of over-fitting at the speaker-level can be seen in the SV-EER results in Tab.~\ref{tab:proposed_systems}.  
    Contrary to usual trends, we notice that the SV-EER for the evaluation partition is almost always lower than that for the development partition.
    We observe this trend for both baseline systems, the pre-trained, fixed system as well as almost all other SASV 2022 submissions, e.g.~\cite{Zhang2022Norm,zhang2022flyspeech,heo2022IRLab}, for which both development and evaluation results are available.
    This consistent finding shows that the evaluation data is somehow \emph{easier} than the development data.
    Curiously, though, results for the jointly-optimised system are inconsistent with this trend;
    it over-fits to the speakers in the training and/or development data.
    
\begin{figure}[!t]
         \centering
         \includegraphics[scale=0.43]{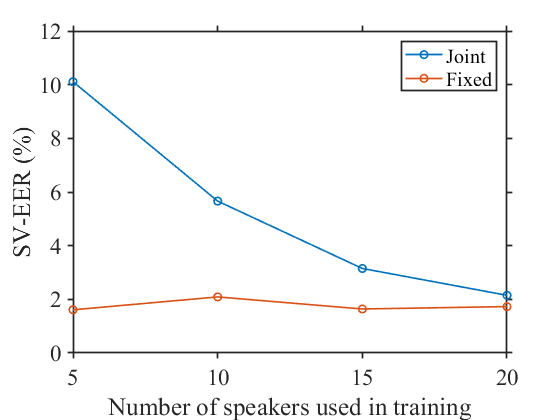}
        \caption{SV-EERs estimated using the development partition for pre-trained, fixed and jointly-optimised systems as a function of the number of speakers in the training partition.}
        \label{fig:sasv-eer}
        \end{figure}

    We designed a set of experiments to test this hypothesis in which we reduced (we could not increase it) the number of speakers in the SASV training data to observe the dependence on SV-EER performance. Results are illustrated in Fig.~\ref{fig:sasv-eer}.
    For the pre-trained, fixed system, the SV-EER is stable at around~2\%. SV-EER results for the jointly-optimised system decrease as the number of speakers increases. Whether or not the use for joint-optimisation of training data containing more speakers will produce SV-EERs below 2\%, or converge to the same level, we cannot say. 
    But, even if it only converges, joint optimisation will still result in improved overall reliability with a lower SASV-EER since it successfully reduces the SPF-EER.
    Our expectation is nonetheless that lower SV-EERs will be achieved through fine tuning since it can mitigate the domain mis-match between VoxCeleb and SASV data.

\section{Conclusions}
\label{sec:conclusion}
    
    We explored the potential of using joint optimisation to improve the reliability of a spoofing-aware speaker verification (SASV) system.
    Results show that, using the SASV 2022 database and protocol, joint optimisation is successful in improving robustness to spoofing attacks by making complementary use of additional information contained within an enrolment utterance.  
    This additional information is either speaker-related or helps to provide a reliable reference, in the form of the enrolment utterances, which are always bona fide.
    Analysis indicates the potential for joint optimisation also to improve speaker verification performance given the availability of training data collected from a greater number of speakers.
    Aside from the collection of new data from a far larger number of speakers, future work should investigate new architectures and loss functions which better exploit the complementarity between spoofing detection and speaker verification systems,  as well as use of enrolment utterances to assist with spoofing detection.
    Other directions include the design of optimisation strategies which better reflect the differences in how information provided by each sub-system is complementary to that provided by the other.
    
    Last, we acknowledge that results reported in this paper are behind those of other competing systems reported in other cited works. 
    We note that most of these employ CM and ASV sub-system ensembles whereas we restricted our work to the consideration of single CM and single ASV classifier sub-systems (no fusion at the CM or ASV sub-system level).  
    This choice was made deliberately to focus on and not complicate unnecessarily the assessment of joint optimisation.
    Even if single-classifier systems are desirable in some practical application scenarios in the interests of computational efficiency and memory footprint, the exploration of, and comparisons to joint optimisation with multiple CM and multiple ASV sub-systems is of interest and another direction for future work.
    
    \section{Acknowledgements}
    
    The first author is supported by the TReSPAsS-ETN project funded by the European Union’s Horizon 2020 research and innovation programme under the Marie Skłodowska-Curie grant agreement No.\ 860813.  The second author is supported by the VoicePersonae project funded by the French Agence Nationale de la Recherche (ANR) and the Japan Science and Technology Agency (JST).

\newpage

\bibliographystyle{IEEEtran}
\bibliography{mybib}

\end{document}